\documentclass[twoside]{article}

\usepackage{PRIMEarxiv}

\usepackage[utf8]{inputenc} 
\usepackage[T1]{fontenc}    
\usepackage{hyperref}       
\usepackage{url}            
\usepackage{booktabs}       
\usepackage{amsfonts}       
\usepackage{nicefrac}       
\usepackage{microtype}      
\usepackage{lipsum}
\usepackage{fancyhdr}       
\usepackage{graphicx}       
\graphicspath{{media/}}     

\usepackage{wasysym}
\usepackage{algorithm2e}
\usepackage{caption}
\usepackage{subcaption}

\title{Towards Using Behavior Trees in Industrial Automation Controllers
}

\author{
  Aleksandr Sidorenko, Mahdi Rezapour, Achim Wagner\\
  Deutsches Forschungszentrum für Künstliche Intelligenz GmbH (DFKI) \\
  Trippstadter Strasse 122, 67663 Kaiserslautern, Germany\\
  \texttt{\{Aleksandr Sidorenko\}aleksandr.sidorenko@dfki.de} \\
   \And
  Martin Ruskowski \\
  Deutsches Forschungszentrum für Künstliche Intelligenz GmbH (DFKI) \\
    Trippstadter Strasse 122, 67663 Kaiserslautern, Germany\\
Institute of machine tools and control systems (WSKL)\\
University of Kaiserslautern-Landau, 67663 Kaiserslautern, Germany\\
}

\begin{document}
\maketitle

\begin{abstract}

  The Industry 4.0 paradigm manifests the shift towards mass customization and cyber-physical production systems (CPPS) and sets new requirements for industrial automation software in terms of modularity, flexibility, and short development cycles of control programs. Though programmable logical controllers (PLCs) have been evolving into versatile and powerful edge devices, which combine real-time guarantees with modern IT and AI technologies, there is a lack of PLC software flexibility and integration between low-level programs and high-level task-oriented control frameworks. Behavior trees (BTs) is a novel framework, which enables rapid design of modular hierarchical control structures and is popular in gaming AI as well as robotic communities. It combines improved modularity with a simple and intuitive design of control logic. This paper proposes an approach for improving the industrial control software design by integrating BTs into PLC programs and separating hardware related functionalities from the coordination logic, which changes frequently during reconfiguration. Several strategies for integration of BTs into PLCs are proposed. The first two integrate BTs with the IEC 61131 based PLCs and are based on the use of the PLCopen Common Behavior Model. The last one utilized event-based BTs and shows the integration with the IEC 61499 based controllers. An application example demonstrates the approach.

  The paper contributes in the following ways. First, we propose a new PLC software design, which improves modularity, supports better separation of concerns, and enables rapid development and reconfiguration of the control software. Second, we show and evaluate the integration of the BT framework into both IEC 61131 and IEC 61499 based PLCs, as well as the integration of the PLCopen function blocks with the external BT library. This leads to better integration of the low-level PLC code and the AI-based task-oriented frameworks. It also improves the skill-based programming approach for PLCs by using BTs for skills composition.

\end{abstract}


\section{Introduction}
\label{sec:Intro}

The Industry 4.0 paradigm manifests the shift towards mass customization and demands a high degree of reconfiguration and flexibility from the production lines. Manufacturers need machines that are both smart and reconfigurable to rapidly meet the changing requirements of the market \cite{morganIndustrySmartReconfigurable2021b}.
One of the challenges in creating such machines lies in designing modular control software, which can be rapidly reconfigured also by non-programmers.

The industrial automation is still dominated by the programmable logical controllers (PLCs), which are programmed in the languages of the IEC 61131 standard and being criticized for not meeting the requirements of the modern distributed control systems \cite{basileImplementationIndustrialAutomation2013}.
Current PLC software is integrated and lacks modularity and reusability. This leads to a low level of separation of concerns in the software development for industrial control. A modern automation engineer must master a wide range of qualifications: starting with the programming on device and communication protocols level and ending with the need of application domain knowledge. With the ever-increasing complexity of machines and processes, this becomes a challenging task, especially for small companies, which cannot afford to hire versatile experts.

Modern PLCs have been evolving into versatile and powerful edge devices, which combine real-time guarantees with IT and AI technologies. Service-oriented architectures and skill-based engineering designs are moving the industry towards next generation cyber-physical production systems (CPPS). Though, there is a gap between AI-driven task-oriented frameworks and low-level real-time PLC-based control. To enable autonomous reconfigurable production modules, a better integration of skill-based software design on the PLC level is required.

\begin{figure*}[!t]
  \includegraphics[width=1.0\textwidth]{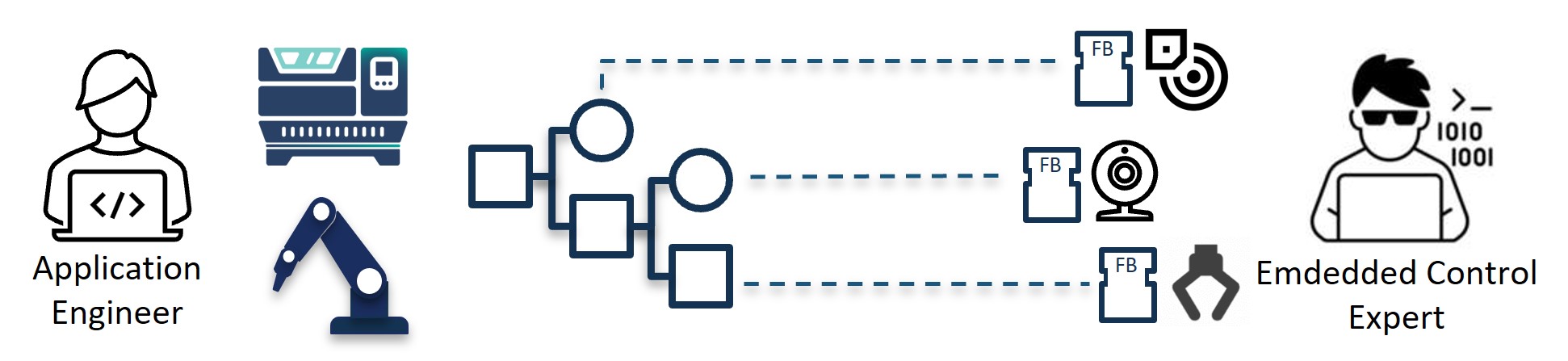}
  \centering
  \caption{BT-based design of industrial control software.}
  \label{fig:BT-based-programming}
\end{figure*}

In this paper, we propose a new approach for designing industrial control software, which is schematically shown in figure \ref{fig:BT-based-programming}. It uses a behavior trees (BTs) supervisory control framework to separate device-specific functions from application logic. This enforces separation of concerns between professional developers of hardware-specific libraries and application engineers. The first ones specialize on the development and support of the device-specific software, the second ones use a low-code framework to rapidly create or reconfigure applications. Furthermore, strict separation between device-specific functions and application logic enables automatic code generation  using AI techniques.

The paper is structured as follows. In section \ref{sec:StoA} we give a short overview of the current methods for developing industrial automation software, as well as some design approaches to improving modularity and flexibility of the PLC programs. Section \ref{sec:Method} presents approaches for integrating behavior trees into PLCs, as well as BT-based design of the PLC programs.  Section \ref{sec:Implementation} shows implementation details and a small application example. Discussion in section \ref{sec:Discussion} concludes the paper.

\section{Related works and technologies}
\label{sec:StoA}

\subsection{Current state of the PLC technology}
\label{subsec:PLC-StoA}

Though introduced more than half a century ago, PLCs are still dominating the industrial automation market, providing reliability and real-time assurances. They have been evolving into versatile edge devices, which combine compliance to strict real-time and safety requirements with flexibility and power of modern IT and AI technologies. Nevertheless, the PLCs are still mostly being programmed using the IEC 61131 standard software model and languages, which are criticized for not meeting the requirements of modern control software. Though the inclusion of object-oriented (OOP) features into the 3rd edition of the standard has improved the modularity and structure of the PLC code, OOP has not provided all the expected benefits, especially in the ares of complexity management and performance \cite{basileImplementationIndustrialAutomation2013}.

Since 1992, a PLCopen association has been striving to improve the quality and efficiency of industrial control software. They encourage using the IEC 61131 standard and their Common Behavior Model to unsure consistency and reusability of the PLC software across different PLC vendors. One of the most used PLCopen specifications and libraries is the PLCopen Motion Control library \footnote{https://plcopen.org/technical-activities/motion-control}, which has harmonized programming for motion control and made it independent of the hardware. In our work, we also align to the PLCopen Common Behavior Model to ensure compatibility of our approach with all the PLCopen complaint software.

To overcome the limitations of the IEC 61131 a new IEC 61499 norm has been proposed, which focuses on the development of distributed automation intelligence. It has significantly improved encapsulation of data and functions in the function blocks and has enabled advanced distributed control systems by introducing an event-based model of computation into PLC-based control. The standard has been widely acknowledged in academia, yet it has not garnered significant attention from the industry. This is largely due to its complexity and difference from the IEC 61131, which was already widely adopted at the time. The situation seems to start changing recently as some big automation vendors developed their commercial products based on the IEC 61499 standard. We show how our approach uses the advantages of the IEC 61499 for designing loosely coupled and distributed control systems.

\subsection{Approaches to designing modular PLC applications.}
\label{subsec:PLC-design}

The increasing complexity of the automation projects and the need to lower their development time and costs require modular and reusable software design. Several approaches to software design focus on improving these qualities.

\subsubsection{Object-oriented design patterns.}

The introduction of OOP features in the third version of the IEC 61131 norm has opened up the possibility of applying a vast body of knowledge gained in the area of OOP design to the PLC software\footnote{https://github.com/0w8States/PLC-Design-Patterns}. By adopting the tested and proven solutions for the frequently occurring problems, one can drastically improve the quality of the PLC code, including modularity, flexibility, and reusability. However, not all the patterns that are commonly used in mainstream programming are applicable to PLCs, provided the specifics of building software for critical systems. Another problem is that, to apply such design patterns, a PLC programmer must be proficient in OOP, which is rarely the case.

\subsubsection{Component-based design.}
\label{subsubsec:CBD}

A component-based design (CBD) is a popular approach in software engineering that aims to improve software modularity and reuse. It views a software component as a “Black-Box” that interacts with other components via predefined interfaces \cite{dai2017applying}. The compliance to the predefined interfaces enables reusability, fast deployment and simple maintenance. The IEC 61499 standard follows the CBD very well. By following the PLCopen guidelines, one can also implement the CBD. The common pattern of CBD is to align the function blocks on the lowest abstraction level with the devices and then compose them into the composite function blocks, which represent more complex, aggregated components, and so on. This creates a natural hierarchy typical for the production environments. Nevertheless, it may lead to tightly coupled software design, as more function blocks are composed one into another.

\subsubsection{Service-oriented architectures on the PLC level.}
\label{subsubsec:SOA}

Service-oriented architecture (SOA) is a paradigm that aims to build loosely coupled, interoperable systems from the software modules or services with clear interfaces. This paradigm is very popular in main stream software engineering and is currently getting attention in the automation domain. \cite{dai2017applying} describes how the SOA can be implemented in the IEC 61499 runtimes. The authors treat a FB as a service and its interface a service contract. The implementation of the SOA with the IEC 61499 FBs is not straightforward and poses some challenges, as the SOA services are normally stateless, and the data is stored apart from logic. The authors also admit that total implementation of SOA on the controller level leads to performance degradation. The service orchestration is made manually as a network of composite FBs, which encapsulate the service FBs.
\cite{basileImplementationIndustrialAutomation2013} proposes a SOA-based design for the IEC 61131 applications, where a new FB-model is used similar to the one from the IEC 61499.
The authors treat these FBs as services and use a Petri-Net supervisor to orchestrate them. This approach enables a good separation of hardware-specific services from the coordination logic. Furthermore, the use of a formal framework, such as Petri-Nets, for service orchestration provides the program's correctness guarantees. The downside is the implementation overhead and design complexity.

In spirit, our work follows the same approaches discussed here, though we try not to introduce additional programming and execution overhead, and keep the design simple and understandable for non-programmers as well.

\subsubsection{Skill-based design.}
\label{subsubsec:Skills}

Skill-based system design can be treated as an evolution of the SOA in the automation domain. Dorofeev in \cite{dorofeevSkillBasedEngineeringIndustrial2020} defines skills as services for control level, stressing their relation to the services from the information technology domain. On the implementation side, there is a clear analogy to the work \cite{basileImplementationIndustrialAutomation2013} presented in the previous section.
In the past years, several skill models and designs have been proposed \cite{dorofeevSkillbasedEngineeringApproach, zimmermannSkillbasedEngineeringControl, volkmannIntegrationFeasibilityContext2021a}  with the common approach of using state machines to define skill's interface. The composition of skills is done either manually by composing the fine-granular skills into bigger ones as in the CBD, or by generating the orchestration code and packing it into the orchestrator-FB \cite{dorofeevGenerationOrchestratorCode}. The first approach leads to tightly coupled, though well-structured design. Generation of the execution code for composite skills from the high-level specifications can be very attractive, but is also computationally expensive and may not be feasible for the real world industrial systems. Another downside is that the generated skills orchestration code may lack transparency for the user.

A software framework that allows intuitively building complex behaviors from the reusable modules without the need for much programming effort is needed to enable frequent and rapid code changes. In \cite{Sidorenko2022} we have proposed to use behavior trees (BTs) framework for skills composition. In this paper, we show how BTs can be used on the PLC level and how they improve the PLC software design.

\begin{table*}[!t]
  \caption{Major BTs nodes types}
  \centering
  \begin{tabular*}{\hsize}{lllll}
    \toprule
    Node type & Symbol & Succeeds & Fails & Running\\
    \midrule
    Sequence &   \(\rightarrow\) &  if all children succeed &  if at least one child fails &  if one child returns Running\\
    Fallback  & \(?\) &  if one child succeeds &  if all children fail &  if one child returns Running\\
    Action &  \(\Box\) &  on completion &  if impossible to complete &  during execution\\
    Condition &  \fullmoon &  if true &  if false &  never\\
    \bottomrule
  \end{tabular*}
  \label{tab:bt-types}
\end{table*}

\section{Methodology}
\label{sec:Method}

\subsection{Behavior trees model.}
\label{subsec:BT-model}

This section briefly and informally introduces a behavior trees model. A detailed introduction to the topic can be found in \cite{colledanchiseBehaviorTreesRobotics2018a}, while \cite{ogrenBehaviorTreesRobot2022a} gives a formal definition of BTs.

\begin{figure}[!h]
  \includegraphics[scale=0.3]{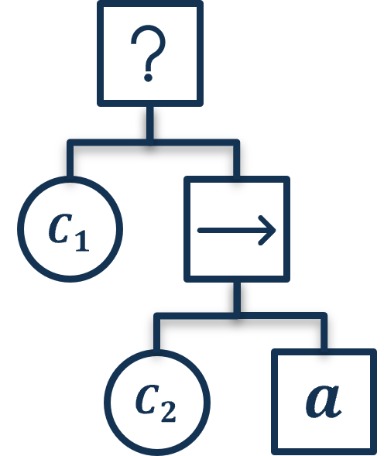}
  \centering
  \caption{Graphical representation of BT.}
  \label{fig:BT-Logo}
\end{figure}

A BT is a control structure that encodes an agent's behavior by mapping the environment states to the agent's actions. It is a directed rooted tree with internal nodes called \emph{control flow nodes} and leaf nodes called \emph{execution nodes}. Conceptually, the execution of BT can be described as follows. The execution starts from the tree root that generates control signals called \emph{ticks} with the given frequency. These signals propagate along the tree and enable the three’s nodes to execute their algorithms. A node is allowed to execute only if it receives a tick signal. After receiving the tick, a child returns immediately \emph{Running} status if it still executes an action, \emph{Success} if it has achieved its goal and \emph{Failure} otherwise. Table \ref{tab:bt-types} summarizes the behavior of the four major BT nodes types: two control flow nodes (sequence and fallback) and two execution nodes (action and condition). The \textbf{\emph{Sequence}} node is used when some actions and conditional checks need to be executed in a sequence, and one action starts only when the previous one succeeds. On the contrary, the \textbf{\emph{Fallback}} node is used when different actions can achieve the same goal. If one action fails, the \textbf{\emph{Fallback}} enables the next one. It fails only if all its children fail.

BTs have a recursive structure, where each subtree can be treated simultaneously as a tree to its child nodes and as an action node to its parent tree. It naturally encodes abstraction levels and priorities. The more abstract functions are located at the top of the tree, while the more detailed ones are at the bottom. Priority levels go from the left to the right under the control flow node. All subtrees share the same basic interface of incoming ticks and returning statuses, which enhances modularity and simplifies composition of control structures.

An example of BT is shown in figure \ref{fig:BT-Logo}. It executes in the following way. When the tick signal comes to the first control flow node (a Fallback node with the question mark in figure \ref{fig:BT-Logo}), it first propagates the signal to the condition node \( C_1 \). If the node \( C_1 \) returns \( Success \) , then the Fallback node and the BT also return \( Success \) to the next upper level. If the node \( C_1 \) fails, then the Fallback node “ticks” its second child, which is a Sequence node. The sequence node executes an action \( a \) only if a condition \( C_2 \) returns \( Success \) . The BT always checks conditions on the path from the root node to the currently running action node. This means that if during the execution of the action \( a \) the condition \( C_1 \) succeeds, then the BT immediately stops the action and returns \( Success \) to the next upper level.

\begin{figure*}[b!]
  \centering
  \begin{subfigure}{.5\textwidth}
    \centering
    \includegraphics[width=.8\linewidth]{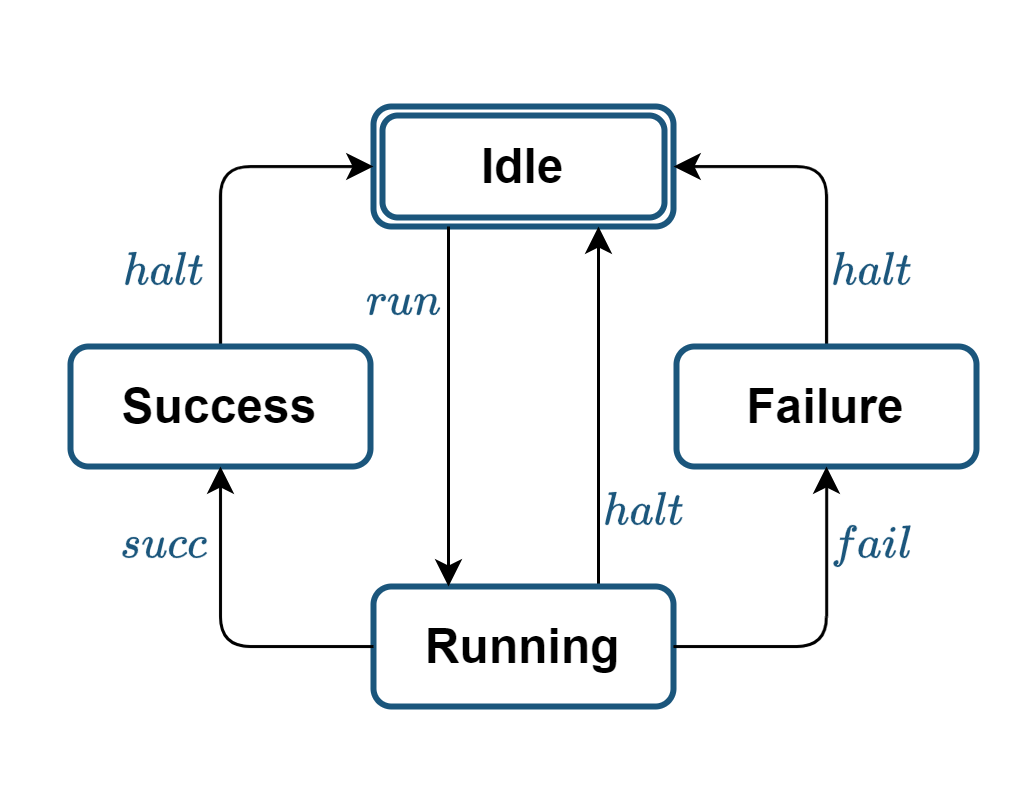}
    \caption{State diagram of BT}
    \label{fig:BT-SM}
  \end{subfigure}%
  \begin{subfigure}{.5\textwidth}
    \centering
    \includegraphics[width=.8\linewidth]{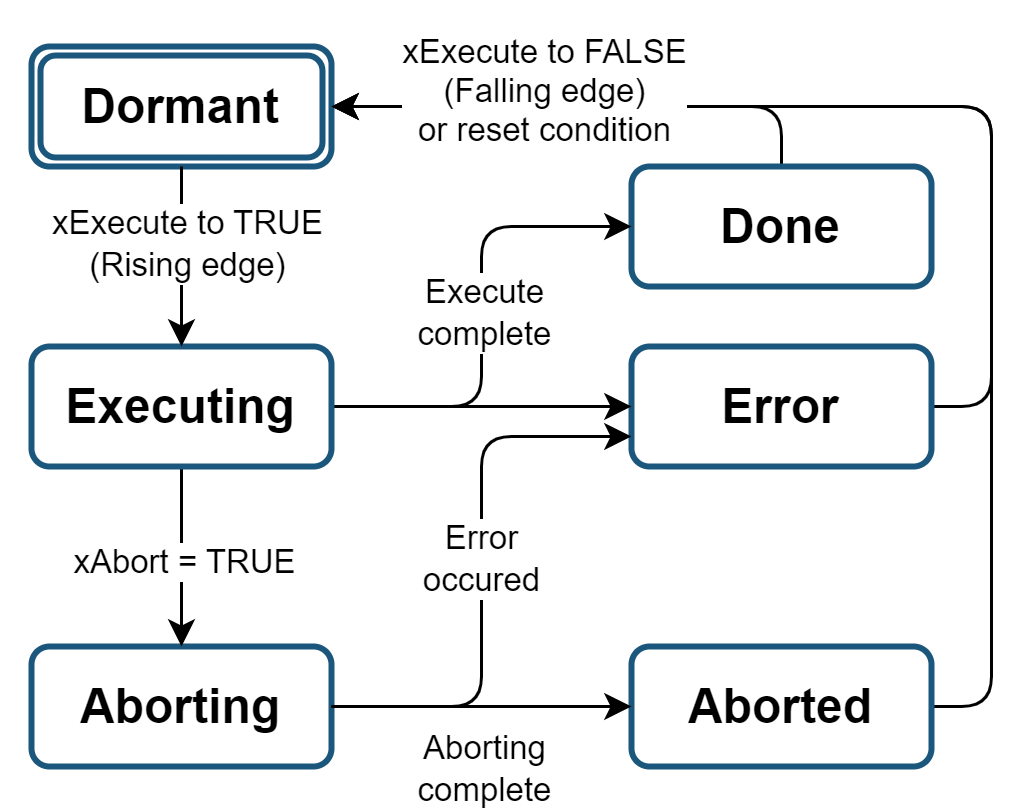}
    \caption{State diagram of ETrigA \cite{noauthor_plcopen_2017}}
    \label{fig:ETrig-SM}
  \end{subfigure}
  \caption{State diagrams of BT and PLCOpen ETrigA FB}
  \label{fig:SMs}
\end{figure*}

The improved modularity of behavior trees is based on 3 major principles. First, it utilizes a recursive nature of the tree structure. Second, it enforces one interface across all the tree's nodes. And third, as feedback about the execution status of its nodes, it uses a small set of symbols \(  \{ S(success), F(failure), R(running) \} \).

A PLCopen Common Behavior Model is a suite of common PLCopen Function Blocks, which was developed by the different PLCopen working groups and described in various specifications and guidelines. By strictly following some key features of the PLCopen specifications, one can achieve a high degree of robustness, usability, and predictability of the PLC software \cite{noauthor_plcopen_2017}. The model provides a number of well-formed, consistent interfaces in terms of inputs and outputs, as well as very similar behaviors explicitly described as state machines, across all the various specifications, which cover major parts of the industrial automation domain.
The FBs defined in the PLCopen specifications are intended to be inherited and extended to suit different needs. This ensures compliance with the guidelines.
The goal is to provide the user with the common modular program structure and a set of simple rules, how to connect modules together.
There are two basic behavior models of the PLCopen FBs: Edge Triggered and Level Controlled FBs. The main difference of the level-controlled behaviors is that they can process new values in each PLC cycle, while the edge-triggered ones need the rising edge of the signal of the \( Execute \)  input. In this work, we show the integration with the Edge Triggered model, but with slight modifications, the same also applies to the Leveled Controlled FBs.

\begin{figure*}[h!]
  \includegraphics[width=1\linewidth]{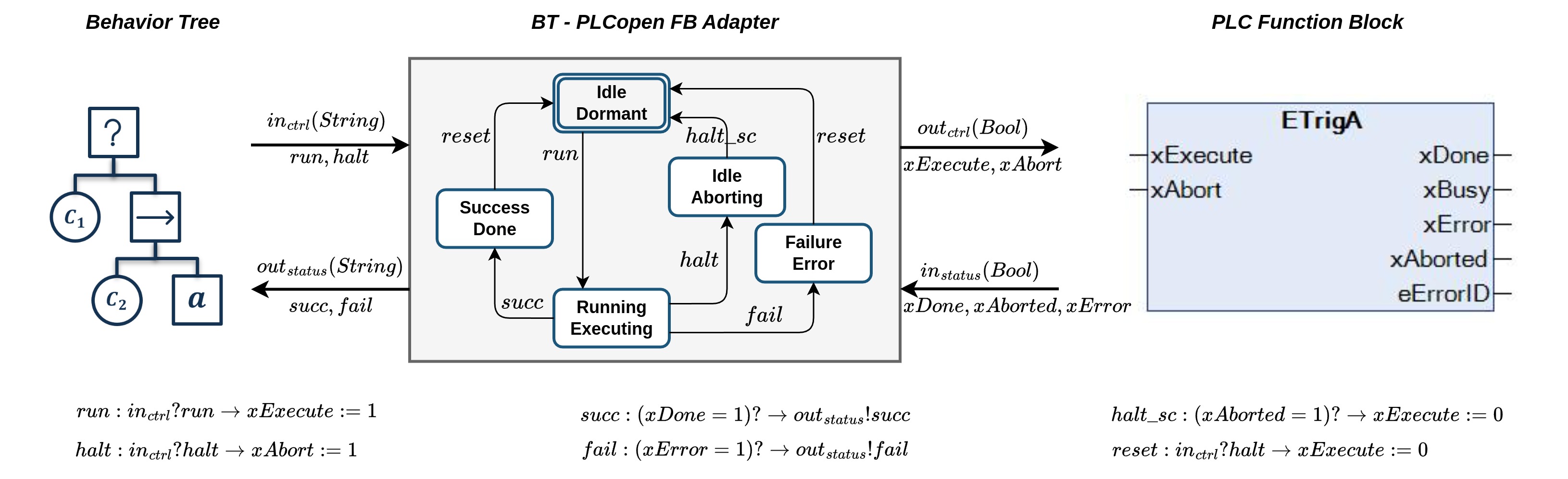}
  \centering
  \caption{Integration of BTs and FBs via adapters.}
  \label{fig:BT-FB-Adapter}
\end{figure*}

\subsection{PLCopen Common Behavior Model.}
\label{subsec:PLCopen-model}

The motivation behind our decision to integrate BTs with the PLCopen is supported by the following considerations. First, PLCopen has a consistent and simple behavior model across all its specifications. It enforces the use of the standard interface with all its function blocks, which is consistent with the BT approach. As will be demonstrated in the subsequent section, the execution semantics of the PLCopen FBs and BTs are very similar, facilitating their integration. Second, PLCopen covers a large part of industrial automation domains, including motion control, safety, and communication. Most major automation vendors have implemented the PLCopen specifications as libraries and integrated the motion control state machine into their servo drives. By integrating BTs with the PLCopen Common Behavior Model, we ensure their compatibility and better chances for BTs to be excepted by the automation community.

\subsection{Integration Strategy.}
\label{subsec:Integration-strategy}

This section discusses several approaches to integration of behavior trees (BTs) and PLC function blocks (FBs).
In the first one, a BT runs as a separate process on the platform different from the PLC. This platform can be, e.g., a Robotic Operating System (ROS). The second and the third integrate BTs directly into PLC code as libraries.

As already mentioned, the behavior models of BTs and PLCopen FBs are very similar. The common pattern can be described as followed. One starts some process with a start signal. The process indicates its execution status with the help of just 3 feedback signals: Running, Success and Failure for the BT, and Executing, Done and Error for the PLCopen FB. The meaning of these feedback signals is the same, so the different naming is irrelevant. Based on this similarity of the two models, we implement the BT model using PLCopen Edge Triggered FB, specifically, ETrigA FB. This is the base Edge Triggered FB with the option to abort the running task with xAbort input. This method is used for the implementation of the BT library for the IEC 61131 PLCs. The implementation details will be shown in the next sections.

To integrate PLC FBs and BTs running on the external platform, we model them as asynchronous processes as in \cite{alur_principles_2015} and \cite{reich_processes_2012}, and use the composition rules for such models. This is because BT's and PLC's runtimes normally run in different threads of control or on the separate hardware devices. The asynchronous model of computation describes such systems very well \cite{alur_principles_2015}. In \cite{Sidorenko2022} we showed how to synchronize asynchronously running BTs using the protocol state machines. A BT as an asynchronous process can be modelled as a simple state machine, shown in Fig. \ref{fig:BT-SM}, which communicates with the other processes asynchronously through signals. The states of the state machine represent the BT return statuses \( \{ Running, Success, Failure\} \). The \( tick \) signal is decomposed into two input signals \( \{ run, halt\} \). The output signals \( \{ succ, fail\} \) indicate the BT's status change.

The ETrigA FB can also be represented as an asynchronous process, which state machine is shown in figure \ref{fig:ETrig-SM}. Its input signals are xExecute to start the process and xAbort to halt it. The output signals \( \{ xDone, xError, xAborted \} \) indicate the state of the running process.
The composition of these two state machines defines an adapter, shown in figure \ref{fig:BT-FB-Adapter}, which integrates the BT and the ETrigA FB.

The IEC 61499 standard has been developed specifically for distributed control applications and uses an event-based execution of the FBs. The asynchronous model can also be applied in this case, where the IEC 61499 FBs can be considered as asynchronous processes, which communicate through the signals, presented as events. We can use the rules of composition for asynchronous processes to implement an event-based variant of BT. The implementation details are shown in section \ref{subsec:IEC61499-BT-Library}.

\begin{figure*}[h]
  \includegraphics[width=1\linewidth]{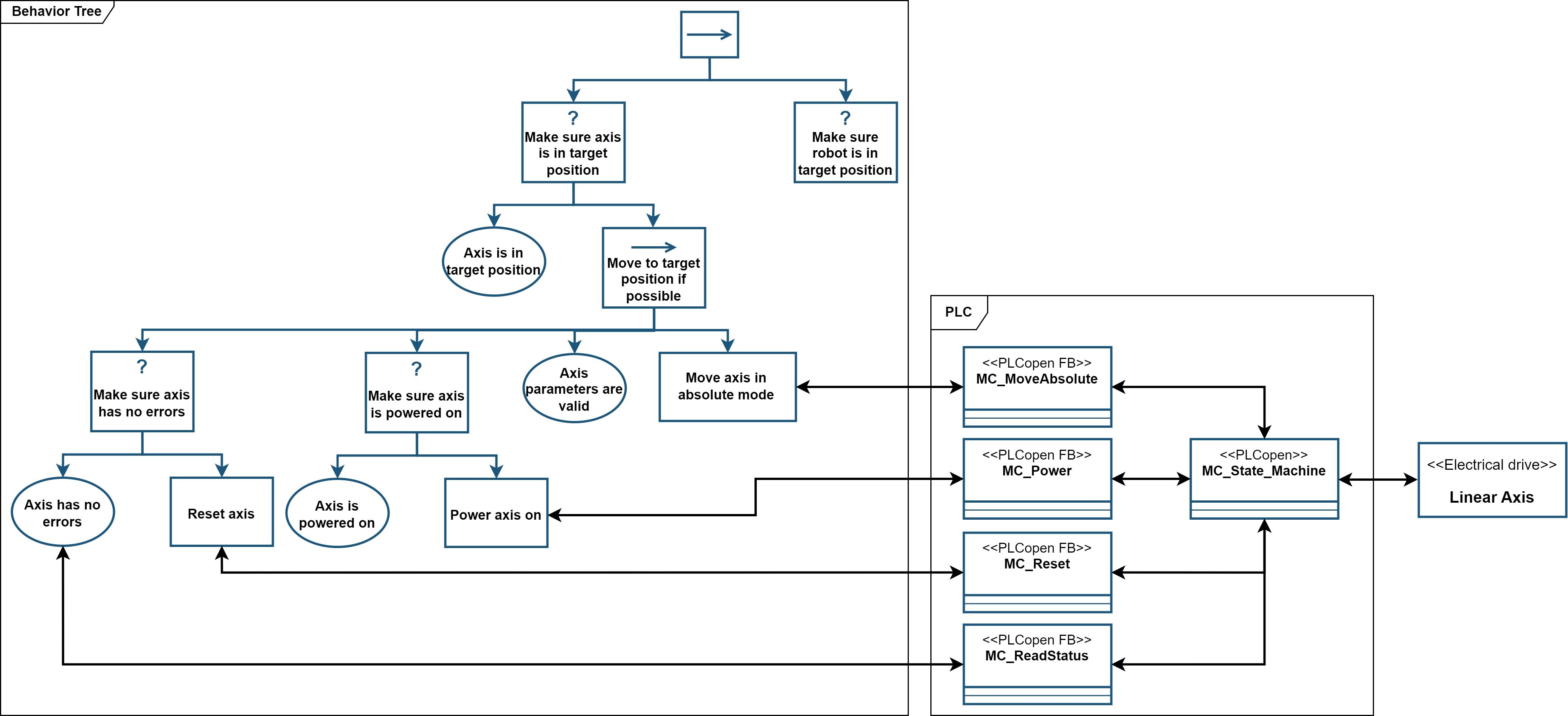}
  \centering
  \caption{BT for controlling PLCOpen liner axis.}
  \label{fig:use-case}
\end{figure*}

\subsection{A BT-based design of the industrial control software}
\label{subsec:Skill-design}

As mentioned in section \ref{sec:Intro} the proposed PLC software design aims to improve modularity and reconfigurability of the PLC programs to ease their development and maintenance.

Figure \ref{fig:BT-based-programming} shows schematically the approach. It follows a well-known in software engineering practice of separating concerns by separating the function related code from the application-specific logic, which coordinates the production process. The functional code normally relates to hardware dependent functionalities, but can also encapsulate some complex algorithms or intellectual property. Such code requires a deep expertise in control and embedded programming. It can be developed by the professionals who specialize in embedded control or supplied by the components producers in the form of well-tested and supported function blocks. The provided FBs must comply with the behavior model of the framework, which is used for designing the application logic. Developing FBs compliant with the PLCopen Common Behavior Model makes them compatibles with the behavior trees, which we use as a composition framework.

An application engineer uses a low/no-code platform based on behavior trees for creating an application. One does not need a programming expertise for using BTs and can focus on the production task at hand and not on the low-level code.
The application engineer must be more of a domain expert than a programmer to successfully accomplish the task. This is so-called bottom up, systems engineering approach for designing the software. An application is built from the ground up by combining well-tested software components into arbitrary complex behaviors.

A top-down approach is also possible, where the application logic is generated from the high-level specifications. The difference of our approach to the other code-generating approaches is that we do not generate all the control code, as, for example, in the supervisory control or in the reactive synthesis. The complexity of the real industrial problems makes such approaches currently infeasible. We assume that the low-level controllers in the form of the PLC FBs already exist and ensure their convergence to the goal. The application code in the form of a behavior tree can then be generated using AI-planning methods.

Using behavior trees to create PLC application logic has several advantages. First, they have a very intuitive design, which is essential for non-programming domain experts, maintainers, and operators. Second, thanks to their modularity, BTs can be easily composed into complex structures. By modifying one part of the tree, one does not normally need to worry about the unwanted interdependencies with the other modules. This speed-ups the reconfiguration of the PLC applications.

\section{Implementation and application example}
\label{sec:Implementation}

\begin{figure*}[!b]
  \includegraphics[width=1.0\textwidth]{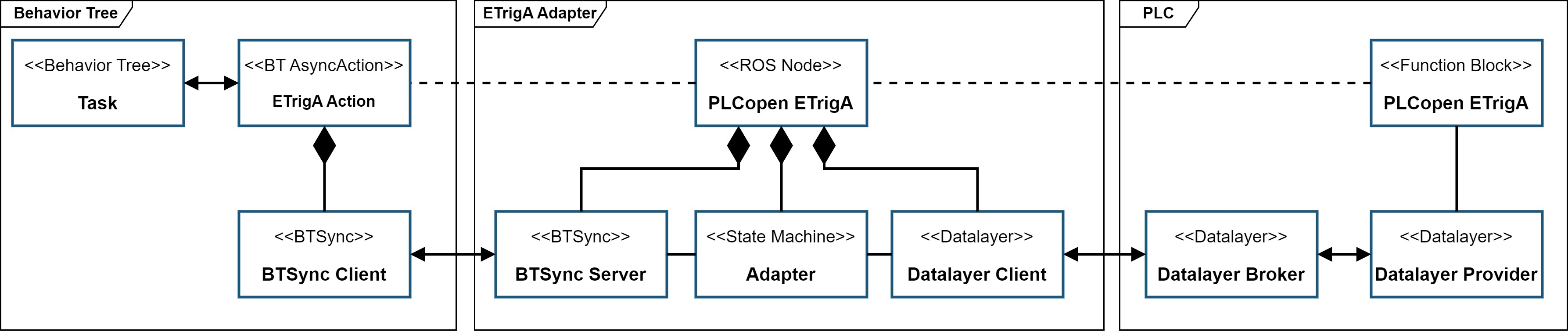}
  \centering
  \caption{Integration of BTs with PLCopen FBs.}
  \label{fig:BT-FBs}
\end{figure*}

To show an application of our approach, we provide the implementation details, as well as a small example. A demonstration setup consists of a robot, which is installed on a linear axis. Both the robot and the axis are controlled by different control systems: the first one is driven by ROS2, the second one — by the PLCopen-compliant drive and IEC 61131 PLC. Figure \ref{fig:use-case} shows a part of the BT, which controls the axis, as this is the focus of our paper. The part that controls the robot is not shown to save space, but it follows the same pattern and is well-presented in literature.

The PLC part consists of the standard PLCopen Motion Control FBs, which interact with the axis state machine. In conventional PLC program design, the application logic is tightly coupled with the PLCopen FB-instances, which makes it difficult to change without changing the complete program. In our design they are not connected to each other and can even be instantiated automatically. The BT actions are synchronized with the motion control FBs for enabling, resetting and moving the axis. The BT conditions get the information from the MC\_ReadStatus FB. The application logic is completely encoded in the BT and can rapidly be changed or generated when needed without touching the low-level functionality.

\subsection{Integration of the PLCopen FBs with the external BT-framework}
\label{subsec:BT-Framework-Integration}

For the implementation of this scenario, we have chosen a ctrlX CORE platform from Bosch Rexroth, as it presents the new generation of automation edge devices and provides a wide variety of possibilities for integration through its Apps concept. The IEC 61131 PLC is one of the ctrlX Apps, which communicates with the other applications through the Data Layer infrastructure. The ROS2 App is used to get access to the robot control system, as well as to run a popular behavior trees engine, BehaviorTree.CPP \footnote{https://github.com/BehaviorTree/BehaviorTree.CPP}.

An integration setup is shown in figure \ref{fig:BT-FBs}. On the PLC side, the PLCopen FBs control the liner axis. A ctrlX Data Layer provider automatically creates an information model of the FBs and provides access to it through a Data Layer broker. The ETrigA adapter, shown in figure \ref{fig:BT-FB-Adapter} is implemented as a ROS node and interacts with the Data Layer on the one side and BT action on the other. The BTSync client of the BT action and BTSync server of the adapter implement the BT synchronization protocol discussed in \cite{Sidorenko2022}.

\begin{figure}[!h]
  \includegraphics[scale=0.16]{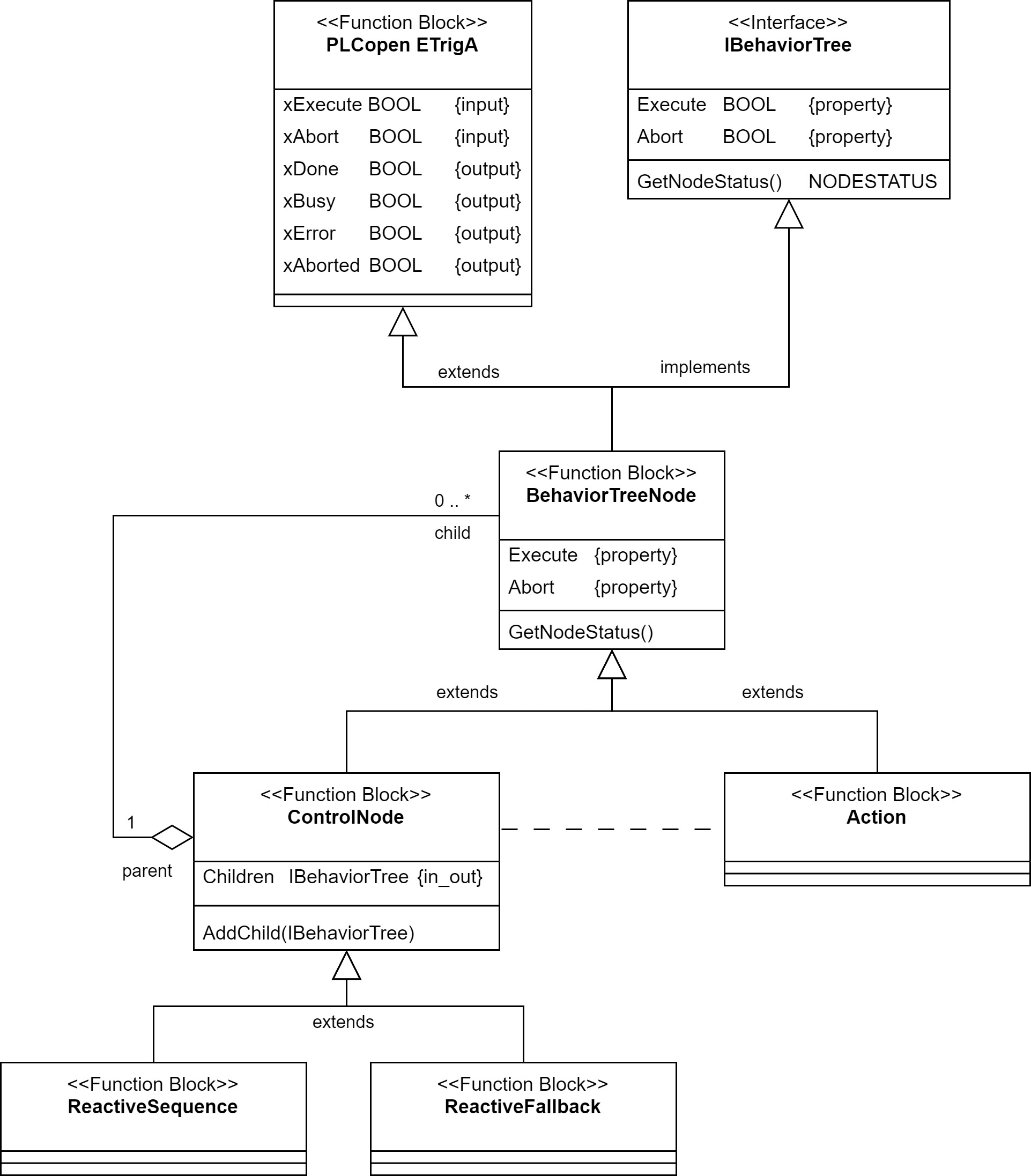}
  \centering
  \caption{Class diagram of BTs for IEC 61131 PLCs.}
  \label{fig:BT4PLC}
\end{figure}

Though the use of an adaptor node brings some overhead, it also fully decouples the PLC side from the BT side. A PLC programmer just needs to create the PLCopen compliant FBs, connect them to the hardware, and provide their addresses in the Data Layer model. On the ROS2 side, the corresponding ROS node takes the address as a parameter, synchronizes with the PLC FB, and starts publishing/subscribing to the specific BTSync topics. A behavior tree designer can use a visual editor, e.g., Groot \footnote{https://www.behaviortree.dev/groot/}, to create a task for the robot and the axis. A special BT-action from the library of actions is used to synchronize with every node, which supports the BT-synchronization protocol from \cite{Sidorenko2022}.

\subsection{BT function blocks for the IEC 61131 PLCs.}
\label{IEC61131-BT-Library}

As discussed in section \ref{subsec:Integration-strategy}, the interfaces and behavior models of the BT execution nodes and the PLCopen FBs are very similar. This can be exploited for creating a BT library for the IEC 61131 based PLCs. The class diagram of the PLC-based BT is shown in figure \ref{fig:BT4PLC}. The \textit{BehaviorTreeNode}, which is the base class of the BT, inherits from the PLCopen ETrigA FB and ensures BT's compatibility with the PLCopen. As all the other BT's nodes inherit from the \textit{BehaviorTreeNode}, the BT is built completely from the ETrigA FBs. Here, the BT tick signal is decomposed into the FB's inputs {xExecute, xAbort} and the BT status consists of the FB's outputs \( \{ xDone, xError, xBusy \} \) .

\RestyleAlgo{ruled}
\SetKwComment{Comment}{/* }{ */}
\begin{algorithm}[hbt!]
  \caption{Pseudocode of a non-blocking Sequence node with \textit{N}-children.}
  \label{alg:bt-sequencer}

  \tcc{ Number of succeded children.}
  $k \gets 0$\;

  \For{ $i \gets 0$ \KwTo $N$}{

    \uIf{$child(i).xBusy$}
    {
      $xBusy \gets True$   \Comment*[r]{ Child is $Running$ }

      \If{$i \geq  k$ }{
        \textbf{break}\;
      }
    }
    \uElseIf{$child(i).xDone$}
    {
      \tcc{Child has succeded.}
      \eIf{$i == N$}
      {
        \tcc{ Last child in sequence.}
        $xDone \gets True$;
      }
      {
        \If{ $ i \geq  k $ }{
          $k \gets k + 1$\;
          \textbf{break}\;
        }

      }
      \textbf{break}\;
    }
    \uElseIf{$child(i).xError$}
    {
      $xError \gets True$  \Comment*[r]{ Child has failed.}
    }
    \uElse{
      $child(i).Execute \gets True$ \Comment*[r]{ child is $Idle$, start it.}
      \textbf{break}\;
    }
  }

\end{algorithm}

The execution model of the classical BT is cyclic and synchronous. This means that in every execution round, the BT's tick signal is propagated across all the tree's nodes according to the rules from the table \ref{tab:bt-types} and is returned as the BT status. This fits in perfectly with the IEC 61131 execution model, which is also cyclic and synchronous.
In the classical implementation of BTs the control flow nodes normally use recursive calls of the \( Tick() \) method on the  $i$-th child \cite{colledanchiseBehaviorTreesRobotics2018a}. To ensure deterministic behavior of the PLCs, recursive and blocking code is not desired. A pseudocode of a non-blocking BT Sequence node is shown in algorithm \ref{alg:bt-sequencer}. It checks the status of each child in the sequence until the first \( Running \) node to be able to react in each cycle.
In general, if the node status check is fast and non-recursive one can use a loop here, though it might be prudent to put a bound on the loop iterations and scan it in several cycles, especially, if a BT is too large. Another possibility would be to decompose the large BT into several smaller ones and run them either in different PLC cyclic tasks or on the separate controllers, and synchronize as shown in \cite{Sidorenko2022}. Because of the synchronous execution model of the IEC 61131 PLCs the order, in which the BT FBs are called, can be important to ensure the deterministic behavior of the BT.

\subsection{Event-based BTs for IEC 61499 controllers.}
\label{subsec:IEC61499-BT-Library}

\begin{figure}[!h]
  \includegraphics[scale=1.0]{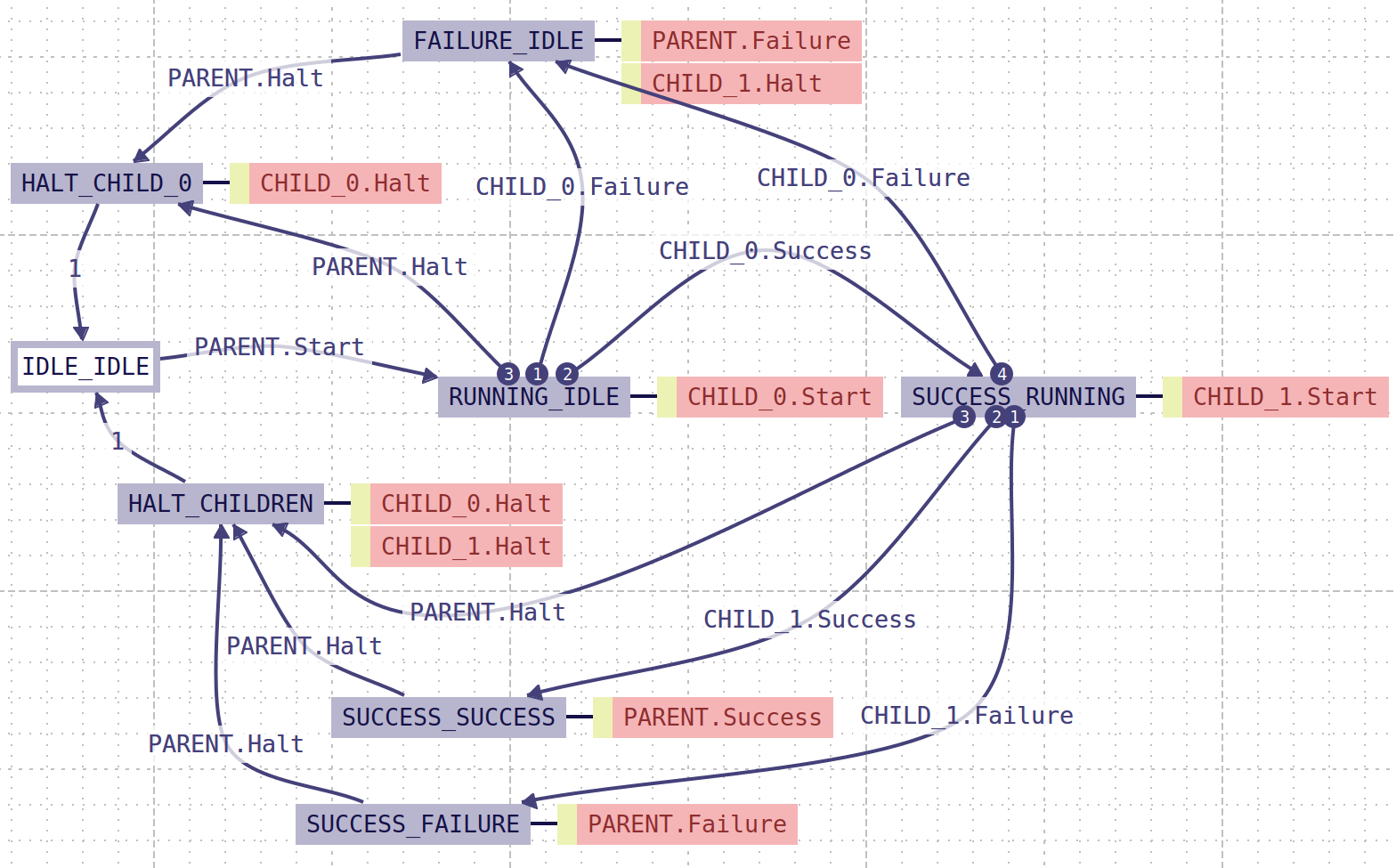}
  \centering
  \caption{Execution control chart (ECC) of the BT Sequence FB in 4DIAC IDE for IEC 61499 controllers.  }
  \label{fig:4diac_sequence}
\end{figure}

\begin{figure*}[!h]
  \includegraphics[width=1.0\linewidth]{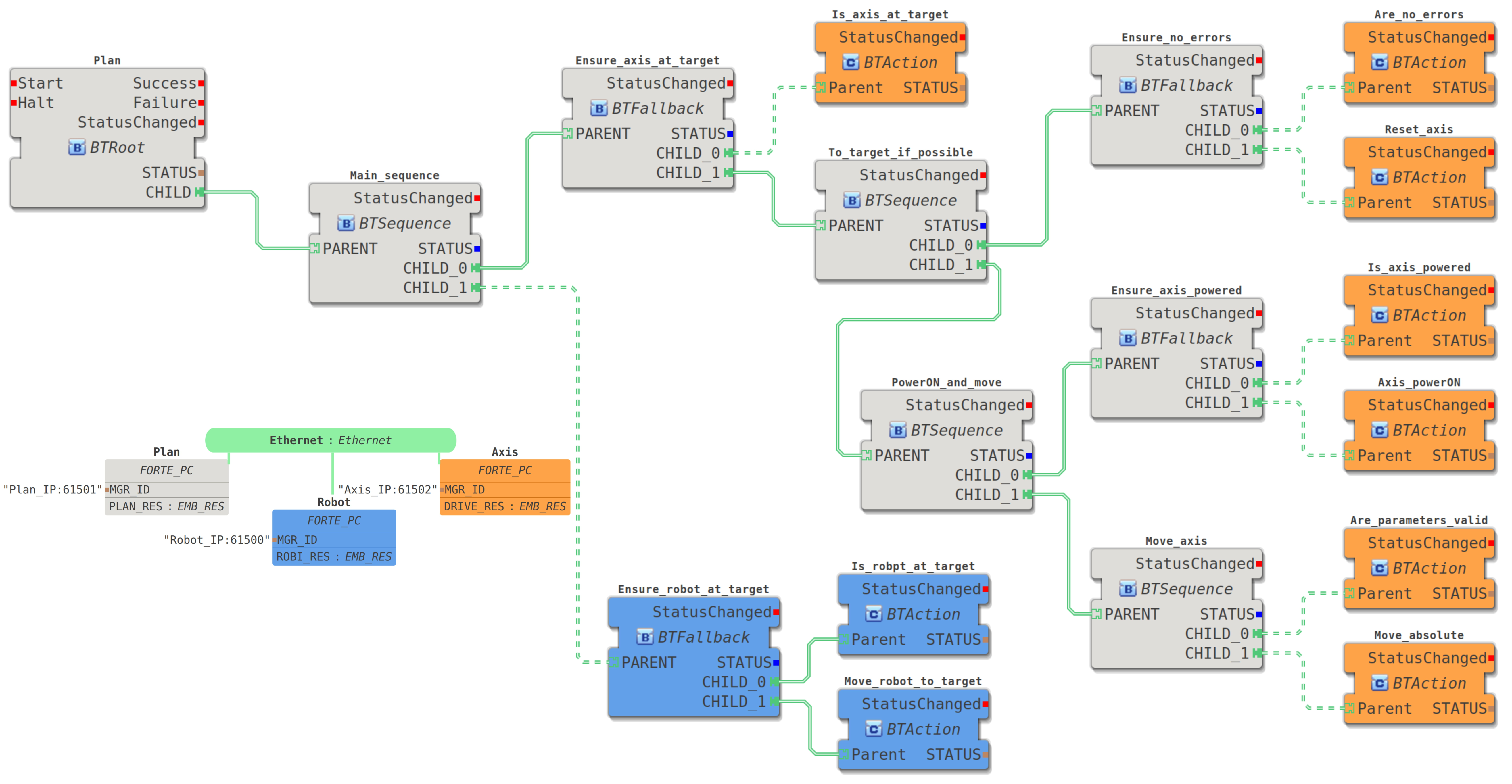}
  \centering
  \caption{Example of BT in 4DIAC IDE for IEC 61499 controllers.}
  \label{fig:4diac-bt}
\end{figure*}

As mentioned in section \ref{subsec:Integration-strategy} the IEC 61499 controllers need the event-based implementation of behavior trees. For this, each BT node is modelled as an asynchronous process, which behavior is defined as the state machine from figure \ref{fig:BT-SM}. The BT execution nodes are implemented as the IEC 61499 basic FBs with the execution control chat (ECC) implementing the BT state machine, and the BT control flow nodes are built by composing the state machines of the execution nodes to enforce the rules from the table \ref{tab:bt-types}. An example of the ECC of the BT Sequence FB for two children is shown in figure \ref{fig:4diac_sequence}. In the ECC, the events \textit{PARENT.x} are used for the communication with the parent of the sequence node, the events \textit{CHILD\_x.x} are for the control of the node's children. The fallback ECC looks almost identical. To make a control flow node for more than two children, one can connect two sequence nodes in the hierarchy as shown in figure \ref{fig:4diac-bt}, though it brings some execution overhead. Another option is to compose together more state machines, each for one child, though this will make the result state machine more complex. However, the synthesis must be done only once, and modern algorithms can handle this task very well. The BT model is extensible by introducing new control flow nodes. They can be synthesized using the same approach of composing BT state machines.

Figure \ref{fig:4diac-bt} shows the IEC 61499 version of the BT from the figure \ref{fig:use-case}. For simplicity, we do not use the PLCopen FBs in this example. As the PLCopen does not focus on the IEC 61499 standard, there are no official implementations of the PLCopen libraries for these types of PLCs, thought, the authors in \cite{sunderAdvancedUsePLCopen2006} showed an approach to implementation of the PLCopen Motion Control library for the IEC 641499 controllers.

As the IEC 61499 controllers have been designed to build distributed control systems, they fit very well for the implementation of distributed BTs, which we described in \cite{Sidorenko2022}. A BT in figure \ref{fig:4diac-bt} is distributed across three runtimes: the orange FBs are running on an autonomous servo drive, as envisioned in \cite{sunderAdvancedUsePLCopen2006}, the blue ones are controlling the robot, and the gray FBs represent the part of the BT, which controls the axis. Here again, the separation of the hardware related functions and the application logic can be seen very well. In fact, as the gray FBs do not have any specific hardware dependencies, they can be automatically generated from the BT high-level specification.

\section{Discussion and Conclusion}
\label{sec:Discussion}

In this work, we have presented an approach for designing industrial control software in a way that enhances modularity and enforces separation of concerns between the device-specific functions, which are programmed as the PLC function blocks (FBs), and the application logic, which is designed using the behavior trees (BTs) framework. We have provided three different approaches for integrating BTs and PLCs.

The first one, which integrates the PLCopen FBs with the external BT library, requires almost no additional effort on the PLC side, except for developing the hardware specific PLCopen-complaint FBs. At the same time, this method uses the adapter on the BT side and is dependent on the communication and integration capabilities of a particular PLC.
Though, modern automation edge devices provide a wide variety of integration possibilities, both device specific, such as Rexroth ctrlX Data Layer, and vendor independent, such as OPC UA communication framework, the integration with the legacy devices may be challenging. The second and the third methods create PLC-based BTs as libraries for two PLC standards: IEC 61131 and IEC 61499. By designing the BT library for the IEC 61131 PLCs, we exploited the similarity between the BT and the PLCopen behavior models, which enabled us to directly use the PLCopen FBs for the BT implementation. As a supervisory control framework, BTs can be a competitive alternative to the IEC 61131–3 Sequential Function Charts (SFC) language, and has several advantages over SFC. BTs provide a higher level of abstraction, they are more modular and better composable into complex behaviors. BTs are also extensible by introducing custom control flow nodes. The tree structure of BTs is arguably more comprehensible and easy to use than the SFC graph, though more empirical studies are required in this domain.

For the IEC 61499 controllers, the event-based version of BT has been implemented, which fits particularly well for controlling distributed applications. It also shows how BTs and state machines complement each other. State machines are more suitable for the development of low-level, hardware related, stateful functionalities. BTs, thanks to their enhanced modularity and compossibility, fit better to the high-level task-oriented supervisory control. In comparison to the AI-planning methods, which are popular at the task level, BTs provide better reactivity to changing environment conditions. In our opinion, BTs can fill the gap between pure AI-planning techniques and hardware related execution of the planned activities.

All tree implementations complement each other and can be used together across various execution platforms and hierarchy levels. The next steps will be to use the approaches described in this work and extend skill-based PLC software design using BTs, as described in \cite{Sidorenko2022}. Furthermore, the integration of planning techniques and PLC-based execution using BTs needs more attention.

\section*{Acknowledgments}
This research has been supported by the European Union’s HORIZON Research and Innovation Program under the grant agreement No 101120218, the project \textbf{HumAIne} \footnote{https://cordis.europa.eu/project/id/101120218}, and conducted for the \textbf{Production Level 4} test bed environment at \textbf{SmartFactoryKL}\footnote{https://www.smartfactory.de/2022-production-level-4-oekosystem/}.

\bibliographystyle{unsrt}
\bibliography{references}

\begin{thebibliography}{10}

\bibitem{morganIndustrySmartReconfigurable2021b}
Jeff Morgan, Mark Halton, Yuansong Qiao, and John~G. Breslin.
\newblock {Industry 4.0} smart reconfigurable manufacturing machines.
\newblock {\em Journal of Manufacturing Systems}, 59:481--506, April 2021.

\bibitem{basileImplementationIndustrialAutomation2013}
Francesco Basile, Pasquale Chiacchio, and Diego Gerbasio.
\newblock On the implementation of industrial automation systems based on
  {PLC}.
\newblock {\em IEEE Transactions on Automation Science and Engineering},
  10(4):990--1003, October 2013.

\bibitem{dai2017applying}
Wenbin Dai, Valeriy Vyatkin, and James~H Christensen.
\newblock Applying {IEC 61499} design paradigms: Object-oriented programming,
  component-based design, and service-oriented architecture.
\newblock In {\em Distributed control applications}, pages 39--68. CRC Press,
  2017.

\bibitem{dorofeevSkillBasedEngineeringIndustrial2020}
Kirill Dorofeev.
\newblock Skill-based engineering in industrial automation domain: Skills
  modeling and orchestration.
\newblock In {\em 2020 IEEE/ACM 42nd International Conference on Software
  Engineering: Companion Proceedings (ICSE-Companion)}, pages 158--161, 2020.

\bibitem{dorofeevSkillbasedEngineeringApproach}
Kirill Dorofeev and Alois Zoitl.
\newblock Skill-based engineering approach using {OPC UA} {Programs}.
\newblock In {\em 2018 IEEE 16th International Conference on Industrial
  Informatics (INDIN)}, pages 1098--1103, 2018.

\bibitem{zimmermannSkillbasedEngineeringControl}
Patrick Zimmermann et~al.
\newblock Skill-based engineering and control on field-device-level with {OPC}
  {UA}.
\newblock In {\em 2019 24th IEEE International Conference on Emerging
  Technologies and Factory Automation (ETFA)}, pages 1101--1108, 2019.

\bibitem{volkmannIntegrationFeasibilityContext2021a}
M.~Volkmann, A.~Sidorenko, A.~Wagner, J.~Hermann, T.~Legler, and M.~Ruskowski.
\newblock Integration of a feasibility and context check into an {OPC} {UA}
  skill.
\newblock {\em IFAC-PapersOnLine}, 54(1):276--281, 2021.

\bibitem{dorofeevGenerationOrchestratorCode}
Kirill Dorofeev et~al.
\newblock Generation of the orchestrator code for skill-based automation
  systems.
\newblock In {\em 2021 26th IEEE International Conference on Emerging
  Technologies and Factory Automation (ETFA)}, pages 1--8, 2021.

\bibitem{Sidorenko2022}
Aleksandr Sidorenko, Jesko Hermann, and Martin Ruskowski.
\newblock Using behavior trees for coordination of skills in modular
  reconfigurable {CPPMs}.
\newblock {\em IEEE International Conference on Emerging Technologies and
  Factory Automation, ETFA}, 2022-September, 2022.

\bibitem{colledanchiseBehaviorTreesRobotics2018a}
Michele Colledanchise and Petter Ögren.
\newblock {\em Behavior {Trees} in {Robotics} and {AI} - {An} {Introduction}}.
\newblock July 2018.

\bibitem{ogrenBehaviorTreesRobot2022a}
Petter Ögren and Christopher~I. Sprague.
\newblock Behavior {Trees} in robot control systems.
\newblock {\em Annual Review of Control, Robotics, and Autonomous Systems},
  5(1):81--107, May 2022.

\bibitem{noauthor_plcopen_2017}
Peter Erning et~al.
\newblock {PLCopen} software creation guidelines: Creating {PLCopen} compliant
  libraries.
\newblock {PLCopen} {Technical} {Document}, PLCopen, May 2017.

\bibitem{alur_principles_2015}
Rajeev Alur.
\newblock {\em Principles of cyber-physical systems}.
\newblock The MIT Press, Cambridge, Massachusetts, 2015.

\bibitem{reich_processes_2012}
Johannes Reich.
\newblock Processes, {Roles} and their interactions.
\newblock {\em Electronic Proceedings in Theoretical Computer Science},
  78:24--38, February 2012.

\bibitem{sunderAdvancedUsePLCopen2006}
C.~Sünder, A.~Zoitl, F.~Mehofer, and B.~Favre-Bulle.
\newblock Advanced use of {PLCopen} motion control library for autonomous servo
  drives in {IEC 61499} based automation and control systems.
\newblock {\em {e \& i} Elektrotechnik und Informationstechnik},
  123(5):191--196, May 2006.

\end{thebibliography}

\end{document}